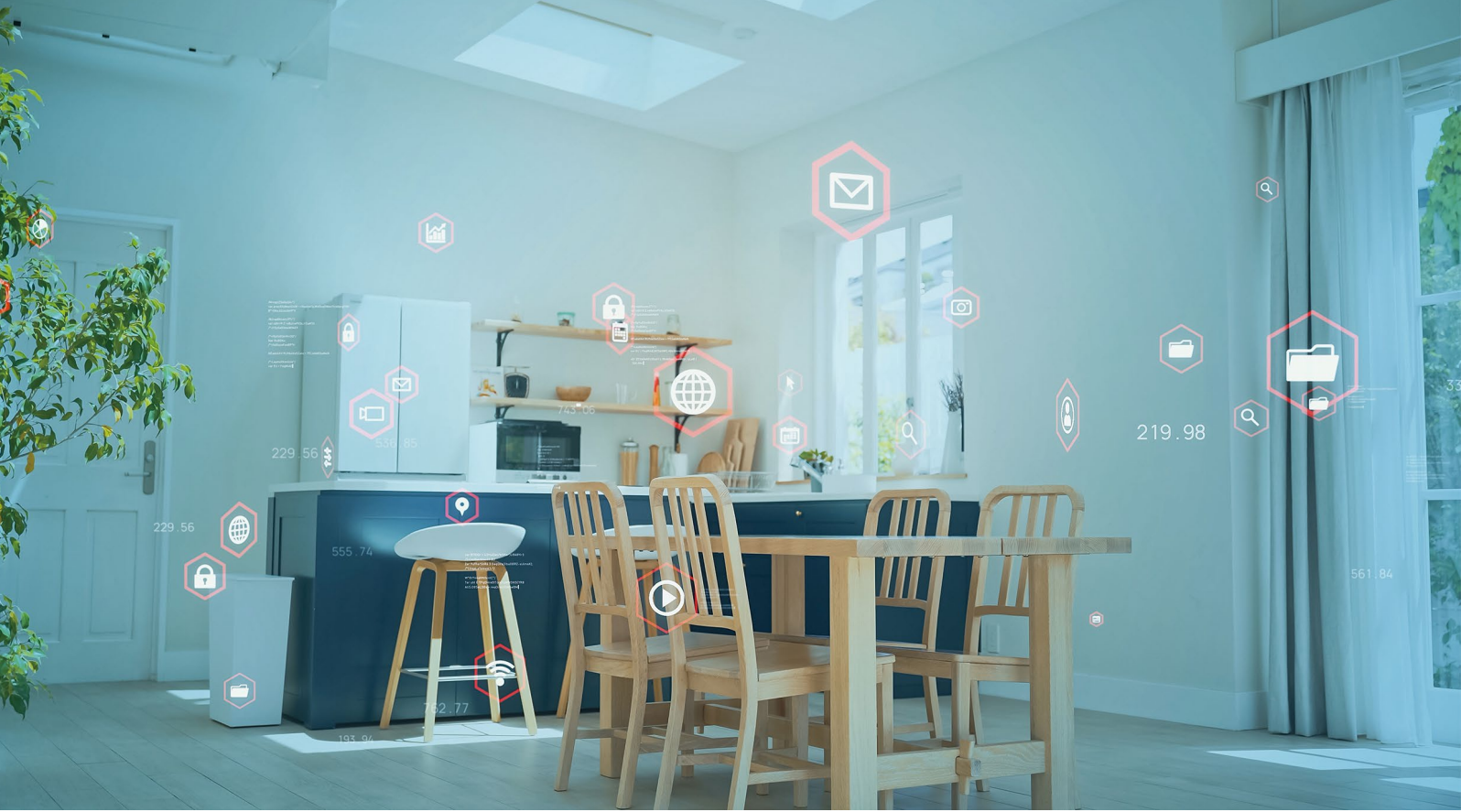

# An Exploration on Integrated Sensing and Communication for the Future Smart Internet of Things

Zhaoxin Chang [1, 4], Fusang Zhang [2], Daqing Zhang [3, 4]

[1] Télécom SudParis

[2] Institute of Software of the Chinese Academy of Sciences

[3] Peking University

[4] Polytechnic Institute of Paris

zhaoxin.chang@telecom-sudparis.eu, fusang@iscas.ac.cn, dqzhang@sei.pku.edu.cn

## Abstract

Internet of Things (IoT) technologies are the foundation of a fully connected world. Currently, IoT devices (or nodes) primarily use dedicated sensors to sense and collect data at large scales, and then transmit the data to target nodes or gateways through wireless communication for further processing and analytics. In recent years, research efforts have been made to explore the feasibility of using wireless communication for sensing (while assiduously improving the transmission performance of wireless signals), in an attempt to achieve integrated sensing and communication (ISAC) for smart IoT of the future. In this paper, we leverage the capabilities of LoRa — a long-range IoT communication technology, to explore the possibility of using LoRa signals for both sensing and communication. Based on LoRa, we propose ISAC designs in two typical scenarios of smart IoT, and verify the feasibility and effectiveness of our designs in soil moisture monitoring and human presence detection.

# 1 Introduction

The rapid development of the Internet of Things (IoT) enriches information exchange applications both between people and things and among things themselves. IoT also enables diversified and intelligent management for these applications by leveraging various sensing devices and wireless communication technologies. In today's applications, however, IoT devices mainly rely on dedicated sensors to sense and collect data, and then transmit the data to target nodes or gateways via wireless communication signals. In recent years, research has demonstrated the sensing capability of wireless communication signals used in the IoT. Analyzing the changes in such wireless signals allows us to sense the status information of people and the environment around IoT devices. With this dual capability of wireless signals, we can explore integrated sensing and communication (ISAC) technologies that enable future IoT devices to support both communication and sensing and consequently replace traditional sensor devices. ISAC technologies create new possibilities for the development of smart IoT. Leveraging wireless signals for sensing not only reduces our reliance on dedicated sensors, but also simplifies the processes of device deployment and maintenance. Utilizing the inherent communication function of IoT devices also makes it possible to transmit sensing data to gateways and the cloud in real time, which is crucial for prompt responses and real-time decision-making.

LoRa is a highly promising IoT technology due to its low power consumption and support for long-range transmission. In urban environments, for instance, LoRa devices can communicate with each other over distances of several kilometers. The long-range capabilities of LoRa make it particularly advantageous in IoT applications and are the reason why it is primarily used in outdoor scenarios. In recent years, researchers have utilized LoRa signals to sense human vital signs in long-range scenarios, demonstrating the unique advantages of LoRa over other types of wireless signals in long-range sensing [1–4]. This paper studies the technical feasibility of applying LoRa signals as an ISAC approach for smart IoT. We explore the sensing and communication requirements in two typical scenarios, and propose corresponding LoRa-based ISAC designs. One design is optimized for outdoor scenarios that require low power consumption more than continuous sensing, whereas the other is tailored for indoor scenarios that require continuous sensing more than low power consumption. For the outdoor and indoor scenarios, we use soil moisture monitoring and human presence detection as separate examples to verify the feasibility of the two designs. Our proposal and verification provide a preliminary solution to the integration of sensing and communication for the future smart IoT.

# 2 Background

## 2.1 LoRa Communication

LoRa signals utilize chirp spread spectrum (CSS) modulation. CSS involves a sequence of chirps whose frequencies increase linearly with time. Each chirp serves as the fundamental data encoding unit for LoRa signals, with all frequencies within a given bandwidth being used to encode the same data. This maximizes the use of the available bandwidth, allowing LoRa signals to overcome attenuation and interference during transmission. As a result, LoRa technologies can support long-range communication over significant distances. As shown in Figure 1, a LoRa data packet consists of the preamble, start frame delimiter (SFD), and data payload. The preamble contains several uplink chirps used for synchronization between the transmitter (a LoRa node) and receiver (a LoRa gateway). The SFD contains several downlink chirps that indicate the start of the payload. By changing the start frequency of chirps, we can realize data encoding and transmission of the payload.

A typical large-scale LoRa network consists of multiple LoRa nodes and LoRa gateways [5]. LoRa nodes collect data from sensors and transmit the data to the gateways. Because LoRa nodes usually transmit small data packets at low frequencies, they mostly remain in sleep mode to conserve power. The LoRa protocol stipulates that the duty cycle for transmitting data from a LoRa node should be kept low, usually not more than 1% [6]. A low duty cycle helps reduce data packet collisions or conflicts among different nodes, ultimately improving the reliability and efficiency of data transmission. In addition, LoRa nodes have a long battery life, lasting for months or even years, making them an ideal option for remote and long-term deployment in wireless sensor networks.

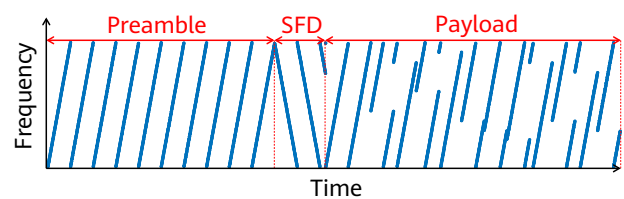

**Figure 1** LoRa data packet structure

## 2.2 LoRa Sensing

Similar to other wireless sensing technologies, LoRa sensing operates by analyzing the relations between received signals and target states. LoRa nodes transmit signals to a LoRa gateway through different paths in an environment. The transmission paths can be straight, reflective, diffractive, or scattering. The signals from these paths are superposed at the receiver to form the received signals. By analyzing the amplitudes and phases of the received signals, we can restore the information of these paths. For instance, we can analyze the changes in the amplitudes and phases of the received signals over time to track the movements of a target person. Additionally, we can analyze the speeds at which the signals are transmitted along straight paths to obtain the properties of the materials that the signals have passed through.

## 3 LoRa-based ISAC Designs

In this section, we propose two LoRa-based designs for two typical scenarios. The first design is intended for outdoor scenarios that require low power consumption more than continuous sensing. To implement sensing in this design, we propose reusing the existing low-rate data packets used for wireless communication. This achieves low-power ISAC because no additional data transmission is needed for sensing. We demonstrate the feasibility of this design by using soil moisture monitoring as an example. The second design focuses on indoor scenarios that require continuous sensing more than low power consumption. We propose adding sensing-dedicated data packets at the transmitter and employing the multi-antenna division operation at the receiver to eliminate amplitude and phase information irrelevant to sensing. This way, we satisfy both communication and continuous sensing requirements, although at the cost of increased power consumption. We have explored the feasibility of this design through human presence detection, where we can determine the walking and still states of people indoors.

## 3.1 ISAC Design by Reusing Communication Data Packets, with Soil Moisture Monitoring as an Example

Soil moisture monitoring plays an important role in smart agriculture. Its implementation is based on the principle that changes in soil moisture levels affect the transmission speeds of wireless signals in soil. When LoRa signals are transmitted over a fixed distance in soil, the varying soil moisture levels cause different degrees of phase change. By measuring the phase change, we can calculate the soil moisture level.

In smart IoT applications, ISAC designs should consider both sensing and communication requirements. To address the sensing aspect, we collect original signals from signal receivers (i.e., gateways) and extract phase information from the signals. When monitoring soil moisture, it is crucial to establish a fixed distance for LoRa signal transmission within the soil and then measure the phase change of signals transmitted over this distance to monitor soil moisture levels. Traditionally, a LoRa transmitter (or node) is usually configured with only one antenna, which can result in non-fixed signal transmission distances in the soil because the distance between the antenna and the soil surface is uncertain. To address this issue, we propose a dual-antenna hardware design for LoRa nodes. This design employs a radio frequency (RF) switch to extend the originally single transmit channel of a LoRa node to two channels, each connected to one antenna. The two antennas are placed in the soil with a known distance between them. LoRa signals are sent from both antennas and then transmitted to a LoRa gateway. At the gateway, the phase difference between the signals sent from the two antennas is measured to estimate the speed of signal transmission in the soil, which is then used to calculate the soil moisture.

In outdoor scenarios, an ISAC design must satisfy the low-power requirements of LoRa communication. To reduce power consumption and extend the battery life of LoRa nodes, the LoRa communication protocol typically allows for a very small duty cycle (specifically, smaller than 1%) for a single LoRa node to transmit signals. While adding data packets for sensing is a direct solution, it will inevitably increase the duty cycle and power consumption. Nonetheless, we have observed that soil moisture monitoring has a low requirement for continuous sensing because the soil moisture often remains stable for a certain period (e.g., one hour). Therefore, when the requirement for low power consumption is high and that for continuous sensing is low, we propose using communication data packets for sensing to avoid the need for adding extra data packets. However, we need to ensure that the newly introduced sensing function does not interfere with the communication function that the data packets are originally intended for. To achieve the dual-antenna design of LoRa nodes, it is necessary to obtain the phase difference between the two antennas placed at different depths in the soil. This can be done by having the two antennas send signals at different times instead of simultaneously. Therefore, during the transmission of a LoRa data packet, we need to switch the antennas in the middle of the transmission process. To minimize the impact of sensing on communication, we should determine an appropriate timestamp for the switching to occur during the transmission of one data packet. As shown in Figure 1, a LoRa data packet includes the preamble and payload, with the data to be transmitted being encoded in the payload. To ensure that the antenna switching will not affect data decoding, we choose to switch the antenna in the preamble part. Figure 2 shows the soil moisture monitoring scenario and its antenna switching process. Before sending a data packet, we reset the RF switch to restore the connection to antenna 1, and start transmitting the preamble part. In the middle of this transmission process, we trigger the RF switch to use antenna 2 for subsequent transmission, which includes the remainder of the preamble that has not yet been transmitted, SFD, and payload. See [7] for details about the design.

We deployed our prototype system in a real-world scenario, as shown in Figure 3b, where LoRa nodes and antennas were placed in the soil. Figure 3c shows the soil moisture measurement precisions achieved at different distances from a LoRa node to the gateway. Our findings indicate that the system's maximum sensing range, i.e., the maximum distance between a LoRa node and the LoRa gateway, can reach up to 100 meters. At this distance, the soil moisture measurement error is about 13%, demonstrating our ISAC system's effectiveness and accuracy in real-world scenarios. We also evaluated the system's communication performance when the sensing function was in operation. The gateway decoded data packets with 100% accuracy, proving that our sensing design does not affect the original communication function.

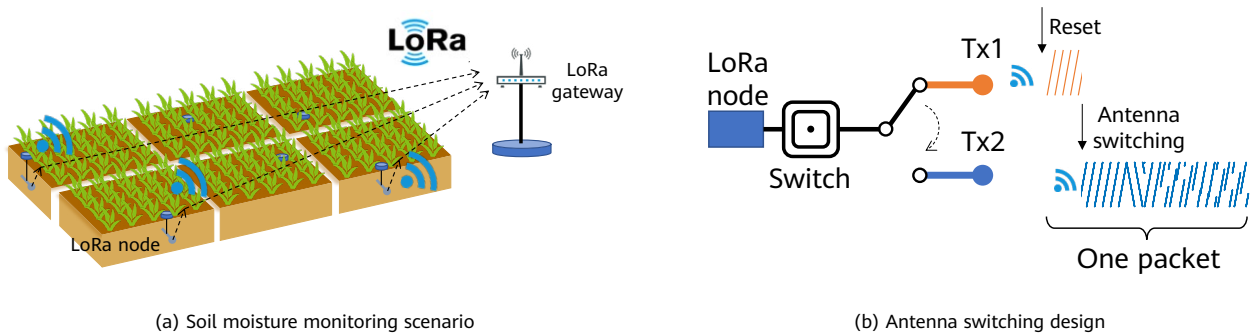

(a) Soil moisture monitoring scenario    (b) Antenna switching design

**Figure 2** Soil moisture monitoring and antenna switching

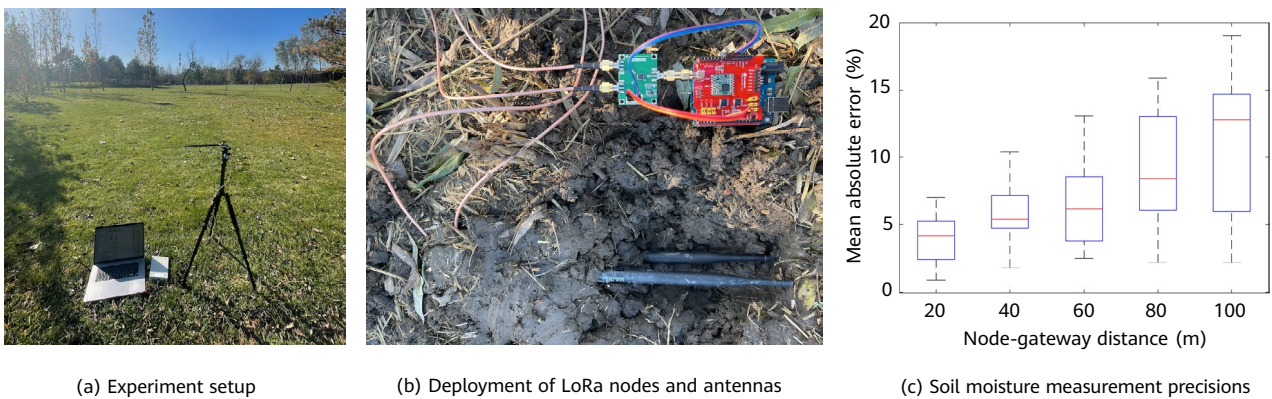

(a) Experiment setup    (b) Deployment of LoRa nodes and antennas    (c) Soil moisture measurement precisions

**Figure 3** Experiment setup and results

ISAC Applications and Demos

## 3.2 ISAC Design for Continuous Sensing, with Human Presence Detection as an Example

Human presence detection is widely used in real-world applications, including smart buildings, security surveillance, and behavior recognition. This detection method primarily relies on the movement of the human body, specifically the walking and still states. The use of wireless signals allows us to detect human presence because the movement of a human body changes the transmission path and distance of the signals reflected by the human body. As a result, the received signals will change with time. By analyzing the changes in the amplitudes and phases in the detection signals, we can determine if human bodies are in the walking or still state in a specific environment. This enables us to achieve human presence detection.

The preceding fundamental principle tells us that continuous sensing, which involves repeatedly transmitting signals, is crucial to ensuring accuracy and preventing missed detections. However, the low duty cycle (typically smaller than 1%) currently stipulated in the LoRa communication protocol does not meet the requirements for continuous sensing. Based on our analysis in the previous section, the limited duty cycle in LoRa communication serves an important purpose in most outdoor application scenarios. It ensures low power consumption, extends battery life, and reduces system maintenance costs. Fortunately, in indoor scenarios, continuous power supply to LoRa nodes is often guaranteed, and there is little requirement for low power consumption. This allows us to develop an ISAC design for continuous sensing in indoor scenarios. Specifically, we enable LoRa nodes to continuously send null data packets that are specially used for sensing, in addition to the original data packets used for communication. Although null packets do not contain any valid data, we can determine human presence by detecting the amplitude and phase changes of these packets at the receiver. Yet there are difficulties in detecting these changes. First, LoRa signals using CSS modulation experience constant frequency variations, resulting in amplitude and phase changes over time. Second, random phase changes can occur due to carrier and sampling frequency offsets in the received signals, because the clocks are not synchronized between LoRa nodes and the gateway. Therefore, in the received signals, we need to eliminate the changes brought by signal modulation and clock asynchronization, and retain only the amplitude and phase changes that are introduced as a result of the varying signal reflection paths. To solve these two issues, we propose configuring two antennas on the LoRa gateway to receive signals. The two antennas are synchronized by the same clock source, resulting in identical phase offsets and modulation characteristics for the signals they receive. However, due to the two antennas being deployed in slightly different spatial locations, the sensing signals they receive may contain minor amplitude and phase changes. By performing a division operation on the two copies of signals received by the two antennas, we can effectively eliminate information irrelevant to sensing. See [1] for details about the principles and signal processing algorithms. Figure 4a illustrates the scenario of human presence detection in a smart building. We deploy a

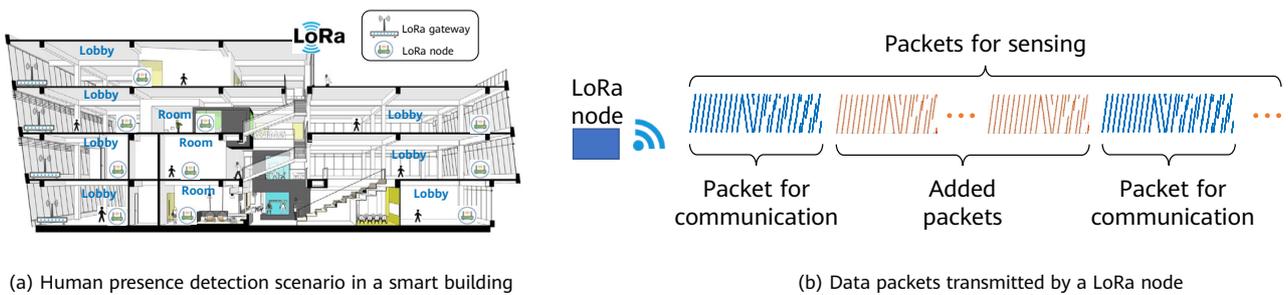

(a) Human presence detection scenario in a smart building     (b) Data packets transmitted by a LoRa node

**Figure 4** Human presence detection and related LoRa data packets

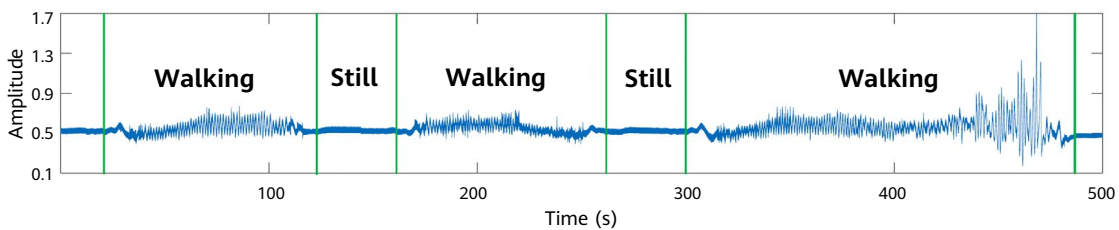

**Figure 5** Signal waveforms for the walking and still states of persons

LoRa gateway on each floor and a LoRa node in every room. This setup allows each LoRa node to detect human presence in its respective room. Figure 4b illustrates the transmission of LoRa data packets based on our ISAC design. In our experiment, we deployed a LoRa node and gateway pair in a 30-meter-long corridor. As shown in Figure 5, the signals obtained through the division between the two antennas exhibit significant fluctuations over time when a person walks in a room and remain constant when a person is still. These signal waveforms enable us to detect human presence and determine their walking and still states.

In the preceding ISAC designs, we used either existing communication packets (in which valid data is encoded) or dedicated null packets to implement sensing. It is noteworthy, however, that the large number of LoRa nodes distributed in a smart building poses a major challenge in eliminating interference between the signals transmitted by various nodes. This can be addressed based on our observation that LoRa nodes can work at different frequencies. When we enable LoRa nodes to transmit signals at different frequencies, a LoRa gateway can receive these signals simultaneously without interference. Alternatively, through proper protocol design, we can ensure that the LoRa nodes working at the same frequency transmit data packets at different time slices to prevent signal conflicts. We have also explored using more antennas on a gateway to implement beamforming for received signals. Beamforming can distinguish signals that come from different spatial orientations, preventing interference and supporting multi-target sensing [2].

## 4 Conclusion

In this paper, we have explored ISAC approaches for the smart IoT of the future, and proposed two LoRa-based designs to meet sensing requirements in two typical scenarios. We have also verified the feasibility of our designs through examples of soil moisture monitoring and human presence detection. We have demonstrated that the wireless signals commonly used in the IoT can be used for sensing the environment and detecting human presence, in addition to their communication purpose. Such ISAC designs based on wireless IoT signals offer several advantages. First, there is no need for extra sensor devices, reducing costs and complexity in system deployment and maintenance when compared to traditional IoT applications. Second, wireless signals can travel long distances and cover large areas, providing both sensing and communication functions with long-range support and wide coverage.